\documentclass[referee]{raa}

\usepackage{graphicx,times}             
\usepackage{natbib}
\usepackage{amssymb,amsmath}

\bibpunct{(}{)}{;}{a}{}{,}
\usepackage[]{hyperref}

\begin{document}

   \title{Revisiting the Evidence for an Intermediate-mass Black Hole in the Center of NGC 6624 with Simulations}

   \volnopage{Vol.0 (20xx) No.0, 000--000}      
   \setcounter{page}{1}          

   \author{Li-Chun Wang
      \inst{1,2}
  \and Yi Xie
      \inst{1}
   }

 \institute{School of Science, Jimei University, Xiamen 361021, Fujian Province, China; {\it xieyi@jmu.edu.cn}\\
 \and
           Physics Experiment Center, Jimei University, Xiamen 361021, Fujian Province, China;\\
\vs\no
   {\small Received~~20xx month day; accepted~~20xx~~month day} }




\abstract{The acceleration of LMXB 4U 1820-30 that derived from its orbital-period derivative $\dot P_{\rm b}$ was supposed to be the evidence for an Intermediate-mass Black Hole (IMBH) in the Galactic globular cluster (GC) NGC 6624. However, we find that the anomalous $\dot P_{\rm b}$ is mainly due to the gravitational wave emission, rather than the acceleration in cluster potential. Using the standard structure models of GCs, we simulate acceleration distributions for pulsars in the central region of the cluster. By fitting the acceleration of J1823-3021A with the simulated distribution profiles (maximum values), it is suggested that an IMBH with mass $M\gtrsim 950^{+550}_{-350}~M_{\odot}$ may reside in the cluster center. We further show that the second period derivative $\ddot P$ of J1823-3021A is probably due to the gravitational perturbation of a nearby star.}


 \authorrunning{WANG \& XIE}            
   \titlerunning{REVISITING THE EVIDENCE FOR AN IMBH IN NGC 6624 WITH SIMULATIONS}  

   \maketitle

\section{Introduction} \label{sec:intro}

Globular clusters are composed of $\sim10^3-10^5$ old stars in regions from tens to hundreds of light years across. They are self-gravitating spherical systems, and become more dense and brighter toward the center. Many studies indicated that IMBHs ($M\sim10^2-10^5 \rm M_\odot$), which are important evolution links between stellar-mass black holes (BHs) and super-massive black holes (SMBHs), should form and reside in dense stellar systems, mainly in the cores of GCs, and also in some dwarf galaxies \citep{2013snpa.confE...2H,1993Natur.364..423S,2001ApJ...562L..19E,2002MNRAS.330..232C,2004ApJ...607...90B,2005ApJ...619L.151B,2006ApJ...641L..21G}. Therefore, some dedicated observations have been performed to search IMBHs in the central regions of the GCs. If confirmed, the evidences would have manifold effects on several basic astrophysical problems, such as the GC dynamics, the origin of ultraluminous X-ray sources and the gravitational wave events \citep{2018ApJ...856...92F}.

There are two schemes that widely used to reveal the presence of IMBHs in GCs: studying the dynamics of the stars and looking for signatures of accretion. However, the existence of IMBHs is still uncertain, and some controversial results were obtained \citep{2004MNRAS.351.1049M,2006ApJS..166..249M,2006ApJ...644L..45P,2006MNRAS.368L..43D,2008ApJ...676.1008N,2010ApJ...710.1063V,2012ApJ...755L...1M,2013ApJ...776..118S,2013A&A...552A..49L,2017Natur.542..203K,2017MNRAS.464.2174B,2018ApJ...862...16T,2019MNRAS.488.5340B,2020MNRAS.499.4646A}. Recently, a luminous X-ray burst from a tidal disruption event in extragalactic star cluster provides that it may contain an IMBH with $M\sim 10^4 M_\odot$ \citep{2018NatAs...2..656L}.

The bulge GC NGC 6624 is one of the most massive and dense clusters located $\sim 1.2~\rm kpc$ away from the Galactic Center and $\sim 7.9~\rm kpc$ away from the Earth \citep{1996AJ....112.1487H}. Due to the cusp signature of optical observations for the radial density profile, it is considered to be a core-collapsed cluster \citep{1995AJ....109..639S}. To date, eleven pulsars have been discovered in the cluster \citep{1994MNRAS.267..125B,2005ApJ...625..951K,2007MNRAS.378..723K,2011Sci...334.1107F,2012ApJ...745..109L,2017MNRAS.468.2114P,2020MNRAS.498..875A,2021MNRAS.504.1407R,2022MNRAS.513.2292A}, and another pulsar J1823-3022 is not yet confirmed as an association with the GC due to its long period and large offset from the cluster center \citep{2022MNRAS.513.2292A}. All of the confirmed pulsars in the cluster have period measurements, and only four of them have period time derivative measurements: PSRs J1823-3021A, B, C, and G. It also contains the low-mass X-ray binary (LMXB) 4U 1820-30, which is the brightest object in X-ray band in the cluster, with extremely short orbital period and an anomalously larger orbital period derivative \citep{1993A&A...279L..21V,2001ApJ...563..934C,2014ApJ...795..116P}.

Analyzing long-term timing of 4U 1820-30 with unprecedented accuracy, \cite{2014ApJ...795..116P} measured the orbital period and its first derivative, $P=(685.01197\pm0.00003)~\rm s$ and $\dot P=(-1.15\pm0.06)\times10^{-12}~\rm{s~s^{-1}}$, which yields $\dot P/P=(-5.3\pm 0.3)\times 10^{-8}~ \rm{yr}^{-1}$. The significant negative $\dot P$ was explained as there is very large amounts of dark remnants contained in the center of NGC 6624. \cite{2017MNRAS.468.2114P,2017MNRAS.471.1258P} further suggested that the cluster may contain an IMBH with the minimum model-dependent mass $\sim 2\times 10^4~\rm M_\odot$, based on the timing data of 4U 1820-30 and PSR B1820-30A. However, the observed period changes of PSR B1820-30A can also be explicated with a cluster model without an IMBH \citep{2018MNRAS.473.4832G}, the dynamical model has a very high core density $\rho_{c}=7.54^{+34.3}_{-5.56}\times10^7 ~M_{\odot}{\rm pc}^{-3}$. Comparing N-body simulations with the cluster, \cite{2019MNRAS.488.5340B} claimed that the model with an IMBH of $M\gtrsim1000 M_\odot$ are incompatible with the optical observation constraints. Although evidence has been accumulating, there is still no uncontroversial case for the existence of IMBHs in any Galactic GCs.

In this work, we simulate the absolute values of the mean-field acceleration for pulsars in NGC 6624, with published density profile and core radius of the cluster \citep{1996AJ....112.1487H}. We fit the measured accelerations of J1823-3021A with the simulated maximum values. This method can provide a minimum mass of the IMBH in the center of the cluster more precisely. In Section 2 we describe the methods of the simulations, and the results are given in Section 3. Our conclusions and discussions are presented in Section 4.

\section{Methods} \label{sec:methods}

\begin{figure*}[!htbp]
\centering
\includegraphics[width=0.5\textwidth]{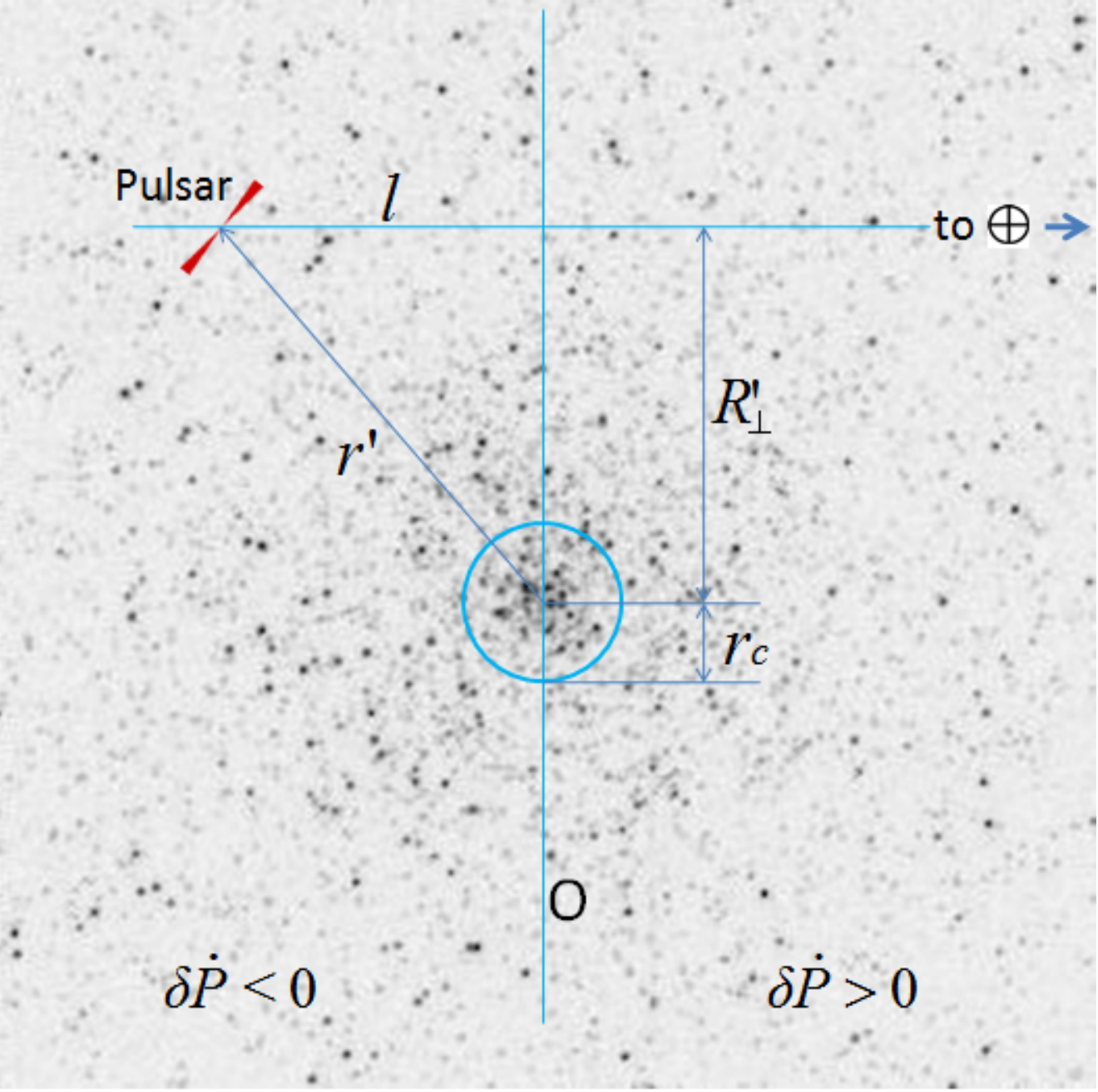}
\caption{Schematic illustration of the location of a pulsar with respect to the centre of the cluster NGC 6624. The optical image of the cluster is taken from the \emph{Hubble Space Telescope} (https://hla.stsci.edu/). The plane of the sky passing through the CoG of the cluster is defined as $O$, $l$ is the line-of-sight distance between the pulsar and plane $O$, $R_{\perp}^{\prime}$ is the projected separation between a pulsar and CoG on the plane of the sky, $r^{\prime}$ is the radial distance of the pulsar from the GC centre, and $r_{c}$ is the core radius of the cluster. When a pulsar lies in the back half of the cluster, the GC potential contributes a negative value on the period of the pulsar, i.e. $\delta\dot P<0$.}\label{fig:0}
\end{figure*}

We first build a coordinate system for the convenient description, as shown in Figure \ref{fig:0}. The plane perpendicular to our line-of-sight and passing through the center of gravity (CoG) of the cluster is defined as $O$. The projected distance of a pulsar from the CoG in plane $O$ as $R_{\perp}^{\prime}$. The line-of-sight distance between the pulsar and plane $O$ as $l$, and $r^{\prime}$ is the radial distance of the pulsar from the GC centre, $r^{\prime}=\sqrt{R_{\perp}^{\prime 2}+l^2}$.

An IMBH strongly affects the spatial distribution of stars around it in the central region. The central IMBH with mass $M_{\rm BH}$ have a radius of influence \citep{2004ApJ...613.1133B},
\begin{equation}\label{influence radius}
r_{i}=\frac{3M_{\rm BH}}{8\pi \rho_{c}r_{c}^2},
\end{equation}
where $r_{c}$ is the core radius of the GC, and $\rho_{c}$ is the core density. At $r_{i}$ and beyond, the density profile are strongly setted by the standard King model \citep{1962AJ.....67..471K},
\begin{equation}\label{King Model}
\rho(r^{\prime})\simeq\rho_{\rm c}[1+(r^{\prime}/r_{\rm c})^2]^{-\frac{3}{2}}.
\end{equation}
Within $r_{i}$, the star distribution is influenced dominantly by the IMBH, the density profile obeys $\rho_{\rm BH}\propto r^{-\alpha}$ \citep{2017ApJ...845..148P}, which is particularly suited for cusp clusters \citep{1986ApJ...305L..61D}. Numerical simulations of clusters with an IMBH in the center found that the power index $\alpha=1.55$ \citep{2004ApJ...613.1143B}.
The total interior mass of stars with contribution at any given radius $r^{\prime}_{\ast}$ can be obtained by radially integrating the density profiles. Multiplying $G/r^{\prime 2}_{\ast}$ yields the acceleration felt at $r^{\prime}_{\ast}$, which reads \citep{2020RAA....20..191X}
\begin{equation}\label{Acceleration}
a(r^{\prime}_{\ast})=\frac{4\pi G}{r^{\prime 2}_{\ast}}[M_{\rm BH}/4\pi+\int_{0}^{r_{i}}r^{\prime 2}\rho_{\rm BH}dr^{\prime}+\int_{r_{i}}^{r^{\prime}_{\ast}}r^{\prime 2}\rho(r^{\prime})dr^{\prime}].
\end{equation}
The mean-field line-of-sight acceleration $a_{l}$ can be obtained via projecting the acceleration $a(r^{\prime}_{\ast})$ along the line-of-sight direction, i.e. $a(r^{\prime}_{\ast})$ is multiplied by a factor of $l/r^{\prime}_{\ast}$. It is apparent that only the stars interior to $r^{\prime}_{\ast}$ are accounted for the acceleration of each pulsar, given the probabilities that the nearest stars contribute to different pulsars (occurring flyby events) at the same time are negligible \citep{2014ApJ...795..116P,2017ApJ...845..148P}.

In order to obtain $a_{l}$ for pulsars at various $R_{\perp}^{\prime}$ of the central region, Monte Carlo simulations are performed for a complete profile of $a_{l}$ distribution. Due to the effect of mass segregation \citep{1987degc.book.....S}, some heavier stellar population like pulsars are centrally concentrated, thus their timing is in favor of giving us insights into the innermost regions of the cluster. It is well confirmed by the locations of reported pulsars, such as these in GCs Terzan 5 and 47 Tucanae. In the circumstances, the observed column number density of the depositing pulsars is given by \citep{1995ApJ...439..191L}:
\begin{equation}\label{density profile}
n(x_{\perp})=n_0(1+x_{\perp}^2)^{\beta/2}.
\end{equation}
in which $n_0$ is the central number density. $x_{\perp}\equiv R_{\perp}^{\prime}/r_{c}$, is the distance away from the CoG in O plane in unit of the core radius of the cluster. $\beta$ is the parameter on the mass segregation, which is preseted to obey a Gaussian distribution, with center value of $-3$ and dispersion of $0.5$ \citep{2019ApJ...884L...9A,2021RAA....21..270W}.

Due to the Doppler effect, the spin period derivatives of pulsars are linked to the line-of-sight accelerations by the equation \citep{1993ASPC...50..141P,2017ApJ...845..148P}
\begin{equation}\label{pdot}
\frac{\dot P}{P}=\frac{\dot P_0}{P_0}+\frac{a_{\rm l}}{c}+\frac{a_{\rm g}}{c}+\frac{a_{\rm s}}{c}+\frac{a_{\rm DM}}{c}
\end{equation}
in which $\dot P_0/P_0$ is the intrinsic spin-down due to the pulsar's magnetic braking, $a_{\rm g}$ is the acceleration from the Galactic potential on a pulsar, $a_{\rm s}$ is the centrifugal acceleration caused by the Shklovskii effect, $a_{\rm DM}$ is the apparent acceleration due to the variations of dispersion measure (DM), $c$ is the speed of light. For NGC 6624, the calculated value of the galactic acceleration is $a_{\rm g}/c=-5.7\times10^{-11}~\rm{yr}^{-1}$ \citep{2014ApJ...795..116P}. The Shklovskii effect for a pulsar in the cluster is $a_{\rm s}/c=\mu^2D/c\approx5.89\times10^{-11}~\rm{yr}^{-1}$ \citep{2017MNRAS.468.2114P}, where $\mu$ is the proper motion of the pulsar, $D$ kpc is the pulsar distance \citep{2003A&A...399..663K}. The apparent acceleration from the stochastic DM error is \citep{2017ApJ...845..148P}
\begin{equation}\label{aDM}
a_{\rm DM}=-6.1\times 10^{-14}(\frac{\Delta t_{\rm DM}}{1 ~\rm{\mu s}})(\frac{10~\rm yr}{T})^2
\end{equation}
in which $\Delta t_{\rm DM}$ is the dispersive delay time, $T$ is the timescale for the DM measurement. For a delay of $\Delta t_{\rm DM}=100~\rm{\mu s}$ and $T=10~\rm yr$, the apparent acceleration is $a_{\rm DM}/c=-6.4\times10^{-15}~\rm{yr}^{-1}$. The line-of-sight acceleration $a_{\rm l}$ depends on the cluster mass within the given radius. At the position of 4U 1820-30, the cluster acceleration is estimated to be $a_{\rm l}/c=1.3\times10^{-9}~\rm{yr}^{-1}$ \citep{2012ApJ...747....4P}. Therefore, $a_{\rm g}$, $a_{\rm s}$, and $a_{\rm DM}$ are all much smaller than $a_{\rm l}$, and have very minor contributions to the measured acceleration. Then, Equation (\ref{pdot}) can be reduced to
\begin{equation}\label{pdot1}
\frac{\dot P}{P}\approx\frac{\dot P_0}{P_0}+\frac{a_{\rm l}}{c}.
\end{equation}
To evaluate the intrinsic spin-down term, we use the median values of the pulsar's $P_0$ and $\dot P_0$ taken from the Australian Telescope National Facility (ATNF) catalog \footnote{https://www.atnf.csiro.au/research/pulsar/psrcat/} \citep{2005AJ....129.1993M}. For millisecond pulsars, the moderate value of $\dot P_0/P_0\sim 10^{-10}~ \rm {yr}^{-1}$. For normal radio pulsar, the value is $\sim 5 \times 10^{-8}~ \rm {yr}^{-1}$. While the magnitude of $a_{\rm l}/c $ could be $\sim 10^{-8}~ \rm {yr}^{-1}$ near the center of the cluster \citep{2001ApJ...563..934C,2018MNRAS.473.4832G}. Thus, $\dot P_0/P_0$ can be neglected only for millisecond pulsars, i.e. $\dot P/P \simeq a_{\rm l}/c$. This relation can be used to test the gravitational potential for the central regions of GCs.

Similar to Equation (\ref{pdot1}), we can express the orbital-period derivative for LMXBs or pulsar binaries as follows:
\begin{equation}\label{pdot orb}
\frac{\dot P_{\rm b}}{P_{\rm b}}=(\frac{\dot P_{\rm b}}{P_{\rm b}})_{\rm GR}+\frac{a_{\rm l}}{c},
\end{equation}
where $(\dot P_{\rm b}/P_{\rm b})_{\rm GR}$ is the orbital decay due to gravitational wave emission. For wide binary systems, the GR term is negligible, the orbital period $P_b$ and orbital period derivatives $\dot P_b$ are very suitable to prob the cluster potential, since $\dot P_b$ is almost completely caused by the cluster potential. Thus Equation (\ref{pdot orb}) may reduce to $\dot P_b/P_b \simeq a_{\rm l}/c$. However, the GR term may be significant for tight binaries, and it has been taken into account in many studies \citep{2016ApJ...818...92M,2019MNRAS.487.1025P}. The X-ray source 4U 1820-30 is in a compact binary system (with a period of $\sim 685$ s), consisting of a white dwarf of $6.89\times 10^{-2}~M_{\odot}$ accreting onto a neutron star of $1.58~M_{\odot}$ \citep{2010ApJ...719.1807G,2021MNRAS.503.5495S}, with an orbit of eccentricity $\lesssim 0.004$ \citep{2007MNRAS.377.1006Z}. Using Equation (8.52) in \cite{2012hpa..book.....L}, we obtained that the expected $(\dot P_{\rm b})_{\rm GR}$ is about $-6.3\times 10^{-12}~ \rm {yr}^{-1}$, and $(\dot P_{\rm b}/P_{\rm b})_{\rm GR}\simeq -2.9\times 10^{-7}~ \rm {yr}^{-1}$. It is apparent that the General Relativity effect is the dominant term. Its absolute value is apparently greater than the value of the observed $\dot P_{\rm b}/P_{\rm b}\simeq -5.3\times 10^{-8}~ \rm {yr}^{-1}$. Part of the reasons may be the accretion process. The secondary (helium dwarf) transfers matter to the accretion disk of the neutron star, which would cause a positive period derivative \citep{1987ApJ...322..842R}. Nevertheless, 4U 1820-30 is currently unsuitable for testing the GC potential.


\section{Results} \label{sec:applications}

There are four pulsars, J1823-3021A, B, C, and G with measured $P$ and $\dot P$, and the LMXB 4U 1820-30 with measured $P_{\rm b}$ and $\dot P_{\rm b}$ in NGC 6624 \footnote{http://www.naic.edu/~pfreire/GCpsr.html}, as shown in Table \ref{Tab:1}. J1823-3021B and J1823-3021C are normal pulsars, $\dot P_0/P_0$ of J1823-3021G is too small (implies the pulsar is located very far away from or very close to the plane $O$, i.e. a very large or small $l$), $\dot P_{\rm b}$ of 4U 1820-30 is contributed dominantly by gravitational wave emission and accretion process, thus only J1823-3021A is suitable for testing the gravitational field of the cluster.

\begin{figure*}[!htbp]
\centering
\includegraphics[width=0.95\textwidth]{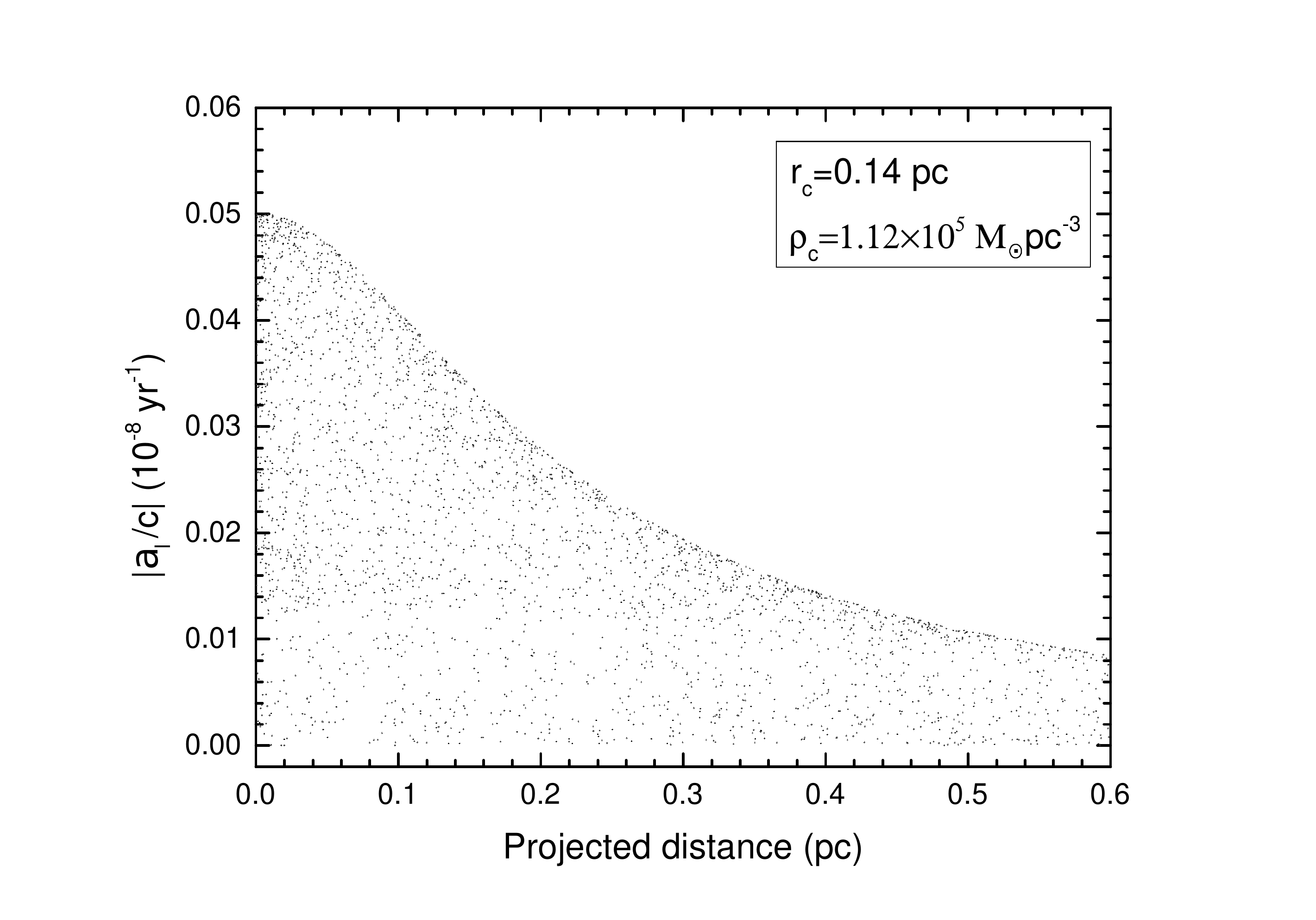}
\caption{The simulated $a_{l}$ due to the gravitational potential of NGC 6624, for pulsars in the innermost region of the GC. The core density $\rho_{c}=1.12\times 10^5~M_{\odot}{\rm pc}^{-3}$ and the core radius $r_{c}=0.14~{\rm pc}$ are taken and no IMBH is considered in the simulation.}\label{fig:1}
\end{figure*}

\begin{table}
	\centering
	\caption{List of Pulsars with Reported Period and its First Derivative in NGC 6624\label{Tab:1}}
	\label{keyp_telescope}
	\begin{tabular}{c|ccccc}
		\hline \hline
		Pulsar & $P$    & $\dot P$               & $\dot P/ P$           & $R_{\perp}^{\prime}$ & Reference\\
       & (ms)   & (10$^{-19}$s~s$^{-1})$ & ($10^{-8}$~yr$^{-1}$) & (pc)&   \\
\hline
J1823-3021A  & 5.44 & 33.85(1) & 1.96 & 0.018(4) &  1,2   \\
J1823-3021B  & 378.6 & 315(3)    & 0.26  & 0.516 &   1,3  \\
J1823-3021C  & 405.9 & 2240(25)  & 1.74 & 0.340  &   3  \\
J1823-3021G  & 6.09 & -0.18(2)  & 0.04  & 0.188  &   4  \\
\hline
4U 1820-30   & $\ast$ & $\ast$ & $\ast$  & 0.050 & 5  \\
		\hline \hline
	\end{tabular}

\emph{References:} (1)\cite{1994MNRAS.267..125B}; (2) \cite{2011Sci...334.1107F}; (3) \cite{2012ApJ...745..109L}; (4) \cite{2021MNRAS.504.1407R}; (5) \cite{2014ApJ...795..116P}.

\end{table}

To obtain the accelerations of pulsars at different locations in NGC 6624, we take the precise values of the central density and the core radius determined from optical observations as $\rho_{c}=1.12\times 10^5~M_{\odot}{\rm pc}^{-3}$ and $r_{c}=0.14~{\rm pc}$, respectively \citep{1996AJ....112.1487H}. We calculate the line-of-sight accelerations ($a_{\rm l}$) caused by the gravitational potential of NGC 6624 with Equations (\ref{influence radius})-(\ref{density profile}). We perform Monte Carlo simulations on the line-of-sight accelerations caused by the gravitational potential of NGC 6624, for pulsars in the innermost region of the GC. The simulated $a_{\rm l}$ distribution is displayed in Figure \ref{fig:1}. One can see that the maximum line-of-sight accelerations, $|a_{\rm l}/c|\lesssim 0.05\times 10^{-8}~{\rm yr}^{-1}$. The simulated results are about two orders of magnitude lower than the reported results of J1823-3021A ($\sim 2\times 10^{-8}~{\rm yr}^{-1}$). Therefore, the mass distribution inferred from the cluster surface brightness cannot explain the acceleration derived from $\dot P/P$ of J1823-3021A. To understand this, an IMBH is contained in the centre or a large amount of dark remnants concentrated in central area, may be the the two most plausible scenarios.

\begin{figure*}[!htbp]
\centering
\includegraphics[width=0.95\textwidth]{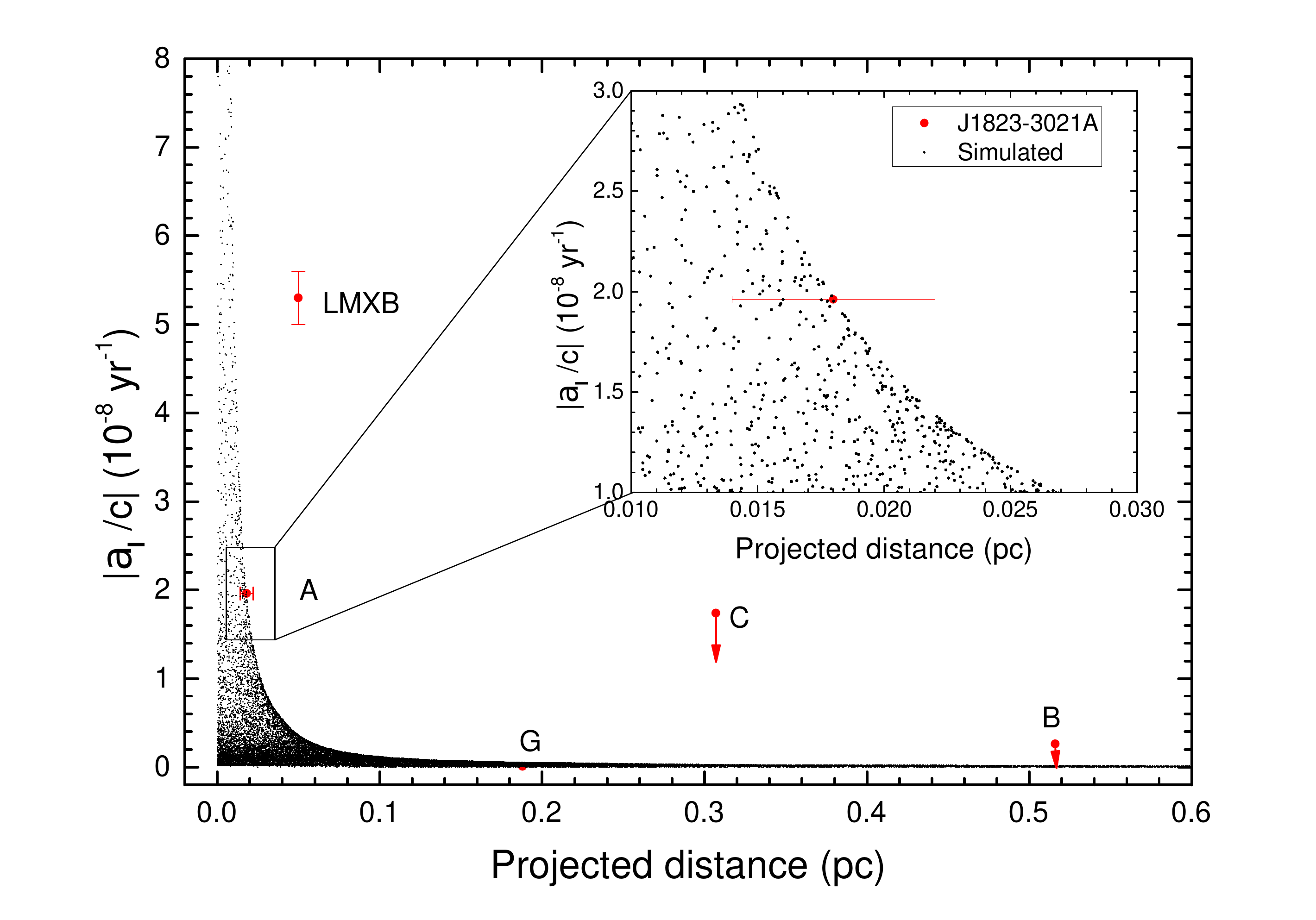}
\caption{The line-of-sight acceleration $a_{l}$ versus the projected radius $R_{\perp}^{\prime}$. The measured accelerations along the line of sight for pulsars J1823-3021A, J1823-3021B, J1823-3021C, J1823-3021G, and LMXB 4U 1820-30 are represented by red points and marked appropriately. The simulated line-of-sight accelerations, which caused by the gravitational potential of NGC 6624 with an additional IMBH of $950~M_{\odot}$, are represented by black circles.}\label{fig:2}
\end{figure*}

For the models, we vary the mass of the central IMBH, while holding all the other cluster parameters constant. We then use the maximum values of the simulated results to fit the acceleration of J1823-3021A. As shown in Figure \ref{fig:2}, the inferred minimum mass of the IMBH is $950^{+550}_{-350}~M_{\odot}$. The uncertainty is obtained by fitting the error bar of the projected distance of the pulsar from the cluster centre. It suggests that an invisible mass of $\sim 950~M_{\odot}$ is embedded in a radius of $\sim 0.02$ pc, providing a mass density of $\gtrsim 1.3\times 10^8~M_{\odot}{\rm pc}^{-3}$. The high-eccentricity orbit of the pulsar reveals the total enclosed mass $M_{t}\gtrsim7.5\times 10^3~M_{\odot}$ \citep{2017MNRAS.468.2114P}, and the averaged mass density $\gtrsim 10^9~M_{\odot}{\rm pc}^{-3}$. Such a high averaged mass density can be interpreted naturally if there is an IMBH in the center, and seems unlikely for the dark remnant scenario.

It is also possible that the unusual $\dot P$ value for J1823-3021A is due to flyby objects in the cluster with its high stellar density, as suggested by \cite{2018MNRAS.473.4832G}. The pulsar will experience a time-varying acceleration due to the passing star, which may produce an anomalous $\ddot P$, as $\ddot P/P=\dot a/c$ \citep{1987MNRAS.225P..51B}. The pulsar acceleration contributed by the close encounter with the nearest star $\sim G M_{t}/b^2$. The average interstellar distance within the radius is $b\sim (3M_s/4\pi \rho_{c})^{1/3}$, where $M_s$ is the mass of the star. The acceleration can be estimated as
\begin{equation}\label{acceleration}
\frac{a}{c}\sim \frac{G M_t}{(3 M_s/4\pi\rho_c)^{2/3}} \simeq 8.2\times 10^{-9}~{\rm yr}^{-1} (\frac{M_t}{100~M_{\odot}})(\frac{M_s}{1~M_{\odot}})^{-2/3}(\frac{\rho_c}{10^5~M_{\odot}{\rm pc}^{-3}})^{2/3}.
\end{equation}
One can get $a/c\sim 8.2\times 10^{-9}~{\rm yr}^{-1}$ for the typical values of the parameters, which is about one third of the observed $\dot P/P$. The time-scale of the encounter can be estimated as $\tau\sim b/\sigma$, where $\sigma$ is the velocity dispersion of stars in the region. Accordingly, $\ddot P/P\sim a/\tau c$ can be expressed as
\begin{equation}\label{pddot}
\frac{\ddot P}{P}\sim \frac{4\pi G \sigma}{3 c}\frac{M_t}{M_s} \rho_c \simeq 1.4\times 10^{-26}~\rm s^{-2} (\frac{\sigma}{10~\rm{km~s}^{-1}})(\frac{M_t}{100~M_{\odot}})(\frac{M_s}{1~M_{\odot}})^{-1}(\frac{\rho_c}{10^5~M_{\odot}{\rm pc}^{-3}}).
\end{equation}
One can get $\ddot P=7.6\times 10^{-29}~\rm s^{-1}$ for the typical values of the parameters. It is sufficient to explain the observed value of the second derivative of the spin period, $\ddot P_{\rm obs}=-1.7\times 10^{-29}~\rm s^{-1}$ \citep{2011Sci...334.1107F}. Since the predicted $\dot P/P$ is much smaller, while the predicted $\ddot P/P$ is greater than four times of the observed values for the same parameters, we conclude that the observed $\dot P/P$ is mainly due to the central IMBH, however, the observed $\ddot P/P$ is probably due to the gravitational perturbation of a flyby object, i.e., it is possible that the pulsar has a close encounter with a nearby star. Thus we confirm that $\dot P$ is dominated by smooth underlying potential, and $\ddot P$ can be used to infer stochastic effects.  The best-fitting dynamical models of \cite{2018MNRAS.473.4832G} obtained a very high central density, $\rho_{c}=7.54^{+34.3}_{-5.56}\times 10^7~M_{\odot}{\rm pc}^{-3}$. It corresponds $\ddot P=2.6\times 10^{-26}~\rm s^{-1}$, and can hardly match with the observed value.

\section{Discussions and conclusions} \label{sec:conclusions}

Theoretical models suggested that IMBHs may form in the central regions of GCs, however, they are still elusive to observations. The timing effects of pulsar accelerations may supply unique tests on the cluster gravitational fields. Based on standard structure models of the GCs, we simulated the acceleration distributions of pulsars in the central region of NGC 6624. By fitting the reported $\dot P/P$ of J1823-3021A, we find that an IMBH with $M\gtrsim 950^{+550}_{-350}~M_{\odot}$ may locate at the cluster center. Compared with the previous approach that fitting the reported data to Equation (7) of \cite{2014ApJ...795..116P}, our method effectively avoids its mathematical approximation (up to $\sim 50 \% $ \citep{1993ASPC...50..141P}), as well as the measurement errors on the line-of-sight velocity dispersion $\sigma(R_\perp)$ of giant stars at the pulsar position.

We found that the second period derivative of J1823-3021A is probably due to a close encounter with a nearby star of the cluster, as previously proposed in literature \citep{2014ApJ...795..116P,2018MNRAS.473.4832G}. $\ddot P_{\rm obs}$ agrees well with the expectation of the model with $\rho_{c}\sim 10^5~M_{\odot}{\rm pc}^{-3}$. We thus confirmed that $\ddot P$ can be used to infer gravitational perturbations \citep{1987MNRAS.225P..51B}. We showed that the orbital-period derivative of 4U 1820-30 is dominated by the effects of the gravitational wave emission and accretion process. Thus it is unsuitable to be used to prob the GC potential. J1823-3021C is also analysed by \cite{2014ApJ...795..116P} and \cite{2017MNRAS.468.2114P}, and an IMBH with mass of $6\times 10^4 ~M_{\odot}$ was derived. However, the acceleration derived from its $\dot P/P$ cannot be used directly to estimate the gravitational potential of the cluster \citep{2017MNRAS.471.1258P}, since J1823-3021C is a normal pulsar, whose $\dot P_0/P_0$ term is usually dominating. To find further evidences for the central IMBH of this cluster, more long-term data may require with other analyzing methods/techniques.

\begin{acknowledgements}
We would like to thank the referee for comments that led to a significant improvement of this paper. We appreciate valuable discussions with J.Y.Liao. This work is supported by National Natural Science Foundation of China under grant Nos. 11803009 and 11603009, and by the Natural Science Foundation of Fujian Province under grant Nos. 2018J05006, 2018J01416 and 2016J05013.
\end{acknowledgements}


\begin{thebibliography}{}
\bibliographystyle{aasjournal}

\bibitem[Abbate et al.(2019)]{2019ApJ...884L...9A} Abbate, F., Possenti, A., Colpi, M., et al.\ 2019, \apjl, 884, L9.
\bibitem[Abbate et al.(2020)]{2020MNRAS.498..875A} Abbate, F., Bailes, M., Buchner, S.~J., et al.\ 2020, \mnras, 498, 875. 
\bibitem[Abbate et al.(2022)]{2022MNRAS.513.2292A} Abbate, F., Ridolfi, A., Barr, E.~D., et al.\ 2022, \mnras, 513, 2292. 
\bibitem[Arca Sedda et al.(2018)]{2018MNRAS.479.4652A} Arca Sedda, M., Askar, A., \& Giersz, M.\ 2018, \mnras, 479, 4652.
\bibitem[Aros et al.(2020)]{2020MNRAS.499.4646A} Aros, F.~I., Sippel, A.~C., Mastrobuono-Battisti, A., et al.\ 2020, \mnras, 499, 4646. 
\bibitem[Barth et al.(2004)]{2004ApJ...607...90B} Barth, A.~J., Ho, L.~C., Rutledge, R.~E., et al.\ 2004, \apj, 607, 90. 
\bibitem[Barth et al.(2005)]{2005ApJ...619L.151B} Barth, A.~J., Greene, J.~E., \& Ho, L.~C.\ 2005, \apjl, 619, L151. 
\bibitem[Baumgardt et al.(2004)]{2004ApJ...613.1133B} Baumgardt, H., Makino, J., \& Ebisuzaki, T.\ 2004, \apj, 613, 1133. 
\bibitem[Baumgardt et al.(2004)]{2004ApJ...613.1143B} Baumgardt, H., Makino, J., \& Ebisuzaki, T.\ 2004, \apj, 613, 1143. 

\bibitem[Baumgardt(2017)]{2017MNRAS.464.2174B} Baumgardt, H.\ 2017, \mnras, 464, 2174.
\bibitem[Baumgardt et al.(2019)]{2019MNRAS.488.5340B} Baumgardt, H., He, C., Sweet, S.~M., et al.\ 2019, \mnras, 488, 5340.
\bibitem[Biggs et al.(1994)]{1994MNRAS.267..125B} Biggs, J.~D., Bailes, M., Lyne, A.~G., et al.\ 1994, \mnras, 267, 125. 
\bibitem[Blandford et al.(1987)]{1987MNRAS.225P..51B} Blandford, R.~D., Romani, R.~W., \& Applegate, J.~H.\ 1987, \mnras, 225, 51P. 
\bibitem[Chou \& Grindlay(2001)]{2001ApJ...563..934C} Chou, Y. \& Grindlay, J.~E.\ 2001, \apj, 563, 934.
\bibitem[de Rijcke et al.(2006)]{2006MNRAS.368L..43D} de Rijcke, S., Buyle, P., \& Dejonghe, H.\ 2006, \mnras, 368, L43.
\bibitem[Djorgovski \& King(1986)]{1986ApJ...305L..61D} Djorgovski, S. \& King, I.~R.\ 1986, \apjl, 305, L61. 
\bibitem[Ebisuzaki et al.(2001)]{2001ApJ...562L..19E} Ebisuzaki, T., Makino, J., Tsuru, T.~G., et al.\ 2001, \apjl, 562, L19. 
\bibitem[Fragione et al.(2018)]{2018ApJ...856...92F} Fragione, G., Ginsburg, I., \& Kocsis, B.\ 2018, \apj, 856, 92. 
\bibitem[Freire et al.(2011)]{2011Sci...334.1107F} Freire, P.~C.~C., Abdo, A.~A., Ajello, M., et al.\ 2011, Science, 334, 1107. 

\bibitem[Gieles et al.(2018)]{2018MNRAS.473.4832G} Gieles, M., Balbinot, E., Yaaqib, R.~I.~S.~M., et al.\ 2018, \mnras, 473, 4832. 
\bibitem[Greene \& Ho(2006)]{2006ApJ...641L..21G} Greene, J.~E. \& Ho, L.~C.\ 2006, \apjl, 641, L21. 
\bibitem[G{\"u}ver et al.(2010)]{2010ApJ...719.1807G} G{\"u}ver, T., Wroblewski, P., Camarota, L., et al.\ 2010, \apj, 719, 1807. 

\bibitem[Haiman et al.(2013)]{2013snpa.confE...2H} Haiman, Z., Tanaka, T., Fernandez, R., et al.\ 2013, SnowPAC 2013 - Black Hole Fingerprints: Dynamics, Disruptions and Demographics, 2

\bibitem[Harris(1996)]{1996AJ....112.1487H} Harris, W.~E.\ 1996, \aj, 112, 1487. 
\bibitem[King(1962)]{1962AJ.....67..471K} King, I.\ 1962, \aj, 67, 471.

\bibitem[Knight et al.(2005)]{2005ApJ...625..951K} Knight, H.~S., Bailes, M., Manchester, R.~N., et al.\ 2005, \apj, 625, 951. 
\bibitem[Knight(2007)]{2007MNRAS.378..723K} Knight, H.~S.\ 2007, \mnras, 378, 723. 

\bibitem[K{\i}z{\i}ltan et al.(2017)]{2017Natur.542..203K} K{\i}z{\i}ltan, B., Baumgardt, H., \& Loeb, A.\ 2017, \nat, 542, 203.
\bibitem[Kuulkers et al.(2003)]{2003A&A...399..663K} Kuulkers, E., den Hartog, P.~R., in't Zand, J.~J.~M., et al.\ 2003, \aap, 399, 663. doi:10.1051/0004-6361:20021781
\bibitem[Lin et al.(2018)]{2018NatAs...2..656L} Lin, D., Strader, J., Carrasco, E.~R., et al.\ 2018, Nature Astronomy, 2, 656. 
\bibitem[Lorimer \& Kramer(2012)]{2012hpa..book.....L} Lorimer, D.~R. \& Kramer, M.\ 2012, Handbook of Pulsar Astronomy, by D. R. Lorimer , M. Kramer, Cambridge, UK: Cambridge University Press, 2012
\bibitem[Lugger et al.(1995)]{1995ApJ...439..191L} Lugger, P.~M., Cohn, H.~N., \& Grindlay, J.~E.\ 1995, \apj, 439, 191.
\bibitem[L{\"u}tzgendorf et al.(2013)]{2013A&A...552A..49L} L{\"u}tzgendorf, N., Kissler-Patig, M., Gebhardt, K., et al.\ 2013, \aap, 552, A49.
\bibitem[Lynch et al.(2012)]{2012ApJ...745..109L} Lynch, R.~S., Freire, P.~C.~C., Ransom, S.~M., et al.\ 2012, \apj, 745, 109. 
\bibitem[Maccarone(2004)]{2004MNRAS.351.1049M} Maccarone, T.~J.\ 2004, \mnras, 351, 1049.
\bibitem[Manchester et al.(2005)]{2005AJ....129.1993M} Manchester, R.~N., Hobbs, G.~B., Teoh, A., et al.\ 2005, \aj, 129, 1993. 
\bibitem[Matthews et al.(2016)]{2016ApJ...818...92M} Matthews, A.~M., Nice, D.~J., Fonseca, E., et al.\ 2016, \apj, 818, 92. 
\bibitem[McLaughlin et al.(2006)]{2006ApJS..166..249M} McLaughlin, D.~E., Anderson, J., Meylan, G., et al.\ 2006, \apjs, 166, 249.
\bibitem[Miller \& Hamilton(2002)]{2002MNRAS.330..232C} Miller, M.~C. \& Hamilton, D.~P.\ 2002, \mnras, 330, 232. 
\bibitem[Miller-Jones et al.(2012)]{2012ApJ...755L...1M} Miller-Jones, J.~C.~A., Wrobel, J.~M., Sivakoff, G.~R., et al.\ 2012, \apjl, 755, L1.
\bibitem[Noyola et al.(2008)]{2008ApJ...676.1008N} Noyola, E., Gebhardt, K., \& Bergmann, M.\ 2008, \apj, 676, 1008.
\bibitem[Perera et al.(2017a)]{2017MNRAS.468.2114P} Perera, B.~B.~P., Stappers, B.~W., Lyne, A.~G., et al.\ 2017a, \mnras, 468, 2114.
\bibitem[Perera et al.(2017b)]{2017MNRAS.471.1258P} Perera, B.~B.~P., Stappers, B.~W., Lyne, A.~G., et al.\ 2017b, \mnras, 471, 1258. 
\bibitem[Perera et al.(2019)]{2019MNRAS.487.1025P} Perera, B.~B.~P., Barr, E.~D., Mickaliger, M.~B., et al.\ 2019, \mnras, 487, 1025. 
\bibitem[Peuten et al.(2014)]{2014ApJ...795..116P} Peuten, M., Brockamp, M., K{\"u}pper, A.~H.~W., et al.\ 2014, \apj, 795, 116. 
\bibitem[Phinney(1993)]{1993ASPC...50..141P} Phinney, E.~S.\ 1993, Structure and Dynamics of Globular Clusters, 50, 141
\bibitem[Pooley \& Rappaport(2006)]{2006ApJ...644L..45P} Pooley, D. \& Rappaport, S.\ 2006, \apjl, 644, L45.
\bibitem[Prager et al.(2017)]{2017ApJ...845..148P} Prager, B.~J., Ransom, S.~M., Freire, P.~C.~C., et al.\ 2017, \apj, 845, 148.
\bibitem[Prodan \& Murray(2012)]{2012ApJ...747....4P} Prodan, S. \& Murray, N.\ 2012, \apj, 747, 4. 
\bibitem[Rappaport et al.(1987)]{1987ApJ...322..842R} Rappaport, S., Nelson, L.~A., Ma, C.~P., et al.\ 1987, \apj, 322, 842. 
\bibitem[Ridolfi et al.(2021)]{2021MNRAS.504.1407R} Ridolfi, A., Gautam, T., Freire, P.~C.~C., et al.\ 2021, \mnras, 504, 1407. 
\bibitem[Sigurdsson \& Hernquist(1993)]{1993Natur.364..423S} Sigurdsson, S. \& Hernquist, L.\ 1993, \nat, 364, 423. 
\bibitem[Sosin \& King(1995)]{1995AJ....109..639S} Sosin, C. \& King, I.~R.\ 1995, \aj, 109, 639. 
\bibitem[Suvorov(2021)]{2021MNRAS.503.5495S} Suvorov, A.~G.\ 2021, \mnras, 503, 5495. 
\bibitem[Spitzer(1987)]{1987degc.book.....S} Spitzer, L.\ 1987, Princeton, N.J. : Princeton University Press, c1987.
\bibitem[Sun et al.(2013)]{2013ApJ...776..118S} Sun, M.-Y., Jin, Y.-L., Gu, W.-M., et al.\ 2013, \apj, 776, 118.
\bibitem[Tremou et al.(2018)]{2018ApJ...862...16T} Tremou, E., Strader, J., Chomiuk, L., et al.\ 2018, \apj, 862, 16.
\bibitem[van der Klis et al.(1993)]{1993A&A...279L..21V} van der Klis, M., Hasinger, G., Verbunt, F., et al.\ 1993, \aap, 279, L21
\bibitem[van der Marel \& Anderson(2010)]{2010ApJ...710.1063V} van der Marel, R.~P. \& Anderson, J.\ 2010, \apj, 710, 1063.
\bibitem[Xie \& Wang(2020)]{2020RAA....20..191X} Xie, Y. \& Wang, L.-C.\ 2020, Research in Astronomy and Astrophysics, 20, 191.
\bibitem[Wang \& Xie(2021)]{2021RAA....21..270W} Wang, L.-C. \& Xie, Y.\ 2021, Research in Astronomy and Astrophysics, 21, 270. 
\bibitem[Zdziarski et al.(2007)]{2007MNRAS.377.1006Z} Zdziarski, A.~A., Wen, L., \& Gierli{\'n}ski, M.\ 2007, \mnras, 377, 1006. 

\end{thebibliography}

\end{document}